\documentclass[aps,prd,twocolumn,nofootinbib,superscriptaddress]{revtex4-2}

\usepackage[T1]{fontenc}
\usepackage[utf8]{inputenc}
\usepackage{graphicx}
\usepackage{amsmath,amssymb}
\usepackage{bm}

\usepackage{orcidlink}
\usepackage{hyperref}

\begin{document}

\title{Nuclear Detonations as Probes of Hidden Superluminal Sectors}

\author{Karl Svozil\,\orcidlink{0000-0001-6554-2802}}
\email{karl.svozil@tuwien.ac.at}
\homepage{http://tph.tuwien.ac.at/~svozil}

\affiliation{Institute for Theoretical Physics,
TU Wien,
Wiedner Hauptstrasse 8-10/136,
1040 Vienna,  Austria}

\date{\today}

\begin{abstract}
We propose a highly speculative phenomenological framework in which nuclear detonations and high-energy collisions serve as probes for hidden sectors with effective superluminal propagation. Motivated by analogies between acoustic and electromagnetic phenomena, we stratify the physical description into three layers: a fundamental ``substrate'' layer, hidden-sector fields with extended causal cones, and the emergent Standard Model. We posit that the extreme, macroscopic stress-energy gradients generated by nuclear explosions might excite substrate or hidden modes that remain kinematically inaccessible to standard laboratory probes. This work unifies various exotic proposals---including extra-dimensional shortcuts and trans-metric shockwaves---into a single formalism, discussing the constraints imposed by causality and observation while outlining how such distinct high-energy regimes could complement one another in searching for physics beyond the emergent metric.
\end{abstract}

\maketitle

\section{Introduction}
\label{sec:intro}

Nuclear detonations are among the most energetic terrestrial events achievable in a controlled fashion. Typical yields in the range $10^{13}$--$10^{17}$~J are released over microsecond timescales, implying transient power outputs $P \sim 10^{19}$--$10^{21}$~W and energy densities that can approach $\rho_E \sim 10^{14}$--$10^{15}$~J/m$^3$. During the earliest stages of a detonation, nuclear, electromagnetic, and hydrodynamic processes all operate in a strongly nonlinear regime over macroscopic volumes.

At the opposite extreme, modern particle accelerators create extremely intense but microscopically localized energy-density spikes in individual collisions, reaching femtometer-scale volumes and attosecond-scale durations. While the total energy per collision is tiny compared to a nuclear device, the peak energy densities can exceed those in nuclear detonations by many orders of magnitude. These collisions can thus be viewed as ``mini-shockwaves'' in the stress-energy tensor.

It is natural to ask whether such extreme conditions---macroscopic in nuclear detonations and microscopic in accelerator collisions---might also excite degrees of freedom beyond the Standard Model (SM). In particular, prior speculative work has proposed:
\begin{enumerate}
    \item Hidden-sector fields coupled feebly to the SM stress-energy tensor, potentially propagating with effective speeds exceeding the speed of light $c$ as perceived in our 3+1-dimensional spacetime;
    \item Extra-dimensional bulk fields that can take geometric ``shortcuts'' in higher-dimensional spacetimes, inducing apparent superluminality along the brane;
    \item A deeper, pre-geometric or ``substrate'' layer whose excitations propagate with characteristic velocity $V_{\rm sub} \gg c$, with SM fields and even the spacetime metric emerging as collective, low-energy phenomena of this substrate.
\end{enumerate}

The aim of this work is to organize these ideas into a single stratified framework with unified notation and phenomenology. We emphasize an analogy from condensed matter physics: the relation between acoustic phonons and electromagnetic (EM) phenomena in solids. There, slow, emergent excitations (sound waves) and fast, fundamental excitations (EM fields, electronic degrees of freedom) are strongly coupled, despite their very different characteristic speeds and effective descriptions. This suggests, at least conceptually, how our relativistic world might itself be an emergent sector of a richer, higher-velocity microphysics.

The present paper synthesizes four threads:
\begin{itemize}
    \item A brane/bulk hidden-scalar model in which nuclear detonations couple to a superluminal hidden field via the trace of the SM stress-energy tensor;
    \item A more general hidden-sector treatment including extra-dimensional geodesic shortcuts, modified dispersion relations, and tachyonic modes;
    \item A ``trans-metric shockwave'' picture in which the spacetime metric is an order parameter of a deeper substrate, and nuclear detonations excite substrate waves propagating at $V_{\rm sub} \gg c$;
    \item An extension of the same framework to high-energy accelerator collisions, treated as microscopic ``mini-shockwaves'' that probe complementary coupling structures.
\end{itemize}
We introduce a common set of fields and speeds, clarify the distinct physical layers, and discuss how apparent superluminality can be consistent with an underlying causal theory.

Throughout we work in units where $\hbar = 1$; we keep $c$ explicit when it is conceptually important to distinguish between different characteristic velocities.

\section{Stratified picture and acoustic analogy}
\label{sec:layers}

\subsection{Acoustic--electromagnetic analogy}

In a crystalline solid, acoustic phonons describe collective lattice displacements with typical speeds
\begin{equation}
    v_s \sim 10^3~{\rm m/s} \ll c,
\end{equation}
and obey an effective relativistic-like equation
\begin{equation}
    \left(\frac{1}{v_s^2}\frac{\partial^2}{\partial t^2} - \nabla^2\right)\phi(\mathbf{x},t) = 0,
    \label{eq:acoustic_wave}
\end{equation}
where $\phi$ is a displacement or pressure field. Within the low-energy elastic theory, $v_s$ defines the sound cone and acts as the maximal signal speed.

Electromagnetic waves in vacuum propagate at $c \simeq 3 \times 10^8~{\rm m/s}$, many orders of magnitude faster than $v_s$. Yet there is strong coupling between acoustic and EM phenomena because both are rooted in the same atomic and electronic microstructure. A paradigmatic example is the acousto-optic effect in transparent crystals~\cite{Brillouin1922,Yariv1984}. A strain field $S_{ij}$ modifies the dielectric tensor $\epsilon_{ij}$, leading to a change in refractive index
\begin{equation}
    \Delta n = -\frac{1}{2} n^3 \, p_{ij} \, S_{ij},
    \label{eq:acoustooptic}
\end{equation}
where $p_{ij}$ is the photoelastic tensor. Microscopically,
\begin{equation}
    p_{ijkl} = \frac{\partial \epsilon^{-1}_{ij}}{\partial S_{kl}},
\end{equation}
encoding how lattice strain affects electronic wavefunctions and therefore EM propagation.

From the point of view of an observer who can probe only phonons, the EM sector would look like a hidden, much faster channel whose excitations are excited indirectly by violent mechanical disturbances. Conversely, the emergent acoustic theory ``knows'' nothing about the underlying light speed $c$; its own causal structure is bounded by $v_s$.

This cross-coupling between fast and slow sectors motivates a guiding intuition, encapsulated in the observation that ``if you slam a door, the whole house shakes'':
violent disturbances in one part of a coupled system inevitably propagate through the underlying structure to affect distant or seemingly unrelated components.
 We ask whether nuclear detonations---the most violent ``door slams'' achievable on Earth---might shake deeper structural layers, if they exist, as well.

\subsection{Three layers of description}

Motivated by this analogy and by emergent-gravity ideas~\cite{VolovikBook,LiberatiReview,VisserAnalog}, we propose a stratified picture of fundamental physics with three conceptually distinct layers:

\begin{description}
    \item[Layer~0: Substrate (pre-geometric) layer] \hfill\\
    A hypothesized microscopic substrate with degrees of freedom described by a field (or set of fields) $\Sigma$. Excitations of $\Sigma$ propagate with characteristic speed $V_{\rm sub}$, which is the true microphysical signal speed. The familiar spacetime metric and SM fields are emergent objects built from $\Sigma$.

    \item[Layer~1: Hidden superluminal sector] \hfill\\
    Fields $\Phi$ that propagate in an extended spacetime (for instance, a higher-dimensional bulk) with effective characteristic speed $v_{\rm h}$ satisfying
    \begin{equation}
        c < v_{\rm h} \le V_{\rm sub}.
    \end{equation}
    These fields are more directly coupled to the substrate than SM fields and can mediate apparent superluminal propagation when projected onto the emergent 3+1-dimensional metric. In some realizations, $\Phi$ may itself be a collective excitation of the substrate, but with a wider causal cone than SM excitations.

    \item[Layer~2: Standard Model sector] \hfill\\
    Familiar fields (photons, gluons, etc.) that propagate on an emergent spacetime with metric $g_{\mu\nu}$, obeying Lorentz invariance with invariant speed $c$. All known interactions and observed signals are confined to this layer.
\end{description}

The hierarchy of speeds
\begin{equation}
    c \ll v_{\rm h} \le V_{\rm sub}
\end{equation}
then parallels the condensed-matter analogy: just as sound ($v_s$) is the emergent causal speed for elastic waves while light ($c$) represents faster microphysics in solids, here $c$ is the emergent causal speed for SM fields while $V_{\rm sub}$ represents the deeper microphysical dynamics.

In what follows we adopt the following notational conventions:
\begin{itemize}
    \item $g_{\mu\nu}$: emergent 4D metric governing SM fields (Layer~2), with light speed $c$;
    \item $G_{AB}$: higher-dimensional bulk metric for hidden fields $\Phi$ (Layer~1);
    \item $V_{\rm sub}$: characteristic microphysical velocity associated with $\Sigma$ (Layer~0);
    \item $T^{\mu\nu}$: SM stress-energy tensor on the 4D brane;
    \item $\Phi$: representative hidden-sector field with superluminal effective propagation;
    \item $\Sigma$: substrate order parameter whose expectation value defines $g_{\mu\nu}$.
\end{itemize}

Nuclear detonations and accelerator collisions will be treated as localized, extreme perturbations in $T^{\mu\nu}$ that can source both $\Phi$ and $\Sigma$ excitations, albeit with very different spacetime structures.

\section{Effective couplings from nuclear detonations}
\label{sec:couplings}

\subsection{Nuclear detonations as stress-energy sources}

A nuclear explosion involves roughly $N \sim 10^{25}$ fission or fusion reactions in a macroscopic region, each converting $\sim\!1$--$10$~MeV of binding energy. For a benchmark 20~kt device,
\begin{align}
    E_{\rm total} &\sim 10^{14}~{\rm J}, \\
    \tau &\sim 10^{-6}~{\rm s}, \\
    V_{\rm init} &\sim 1~{\rm m}^3,
\end{align}
yielding an instantaneous power $P \sim 10^{20}$~W and energy density $\rho_E \sim 10^{14}$~J/m$^3$. The resulting stress-energy tensor
\begin{equation}
    T^{\mu\nu}(x) = T^{\mu\nu}_{\rm nuc}(x) + T^{\mu\nu}_{\rm EM}(x) + T^{\mu\nu}_{\rm hydro}(x) + \dots
\end{equation}
varies rapidly in both space and time.

We will typically characterize the source by a coarse-grained trace
\begin{equation}
    T(x) \equiv T^\mu{}_\mu(x),
\end{equation}
which is nonzero due to rest-mass contributions and the QCD trace anomaly. Within QCD,
\begin{equation}
    T^\mu{}_\mu = \frac{\beta(g_s)}{2 g_s} G^a_{\mu\nu} G^{a\mu\nu} + \sum_q m_q \bar{q} q,
\end{equation}
where $g_s$ is the QCD coupling, $G^a_{\mu\nu}$ the gluon field strength, and $m_q$ the quark masses. During a detonation, both terms fluctuate violently.

\subsection{Layer~1: Hidden field sourced by $T$}
\label{subsec:layer1_coupling}

At the hidden-sector level (Layer~1), we posit a scalar field $\Phi$ living in a $(4+1)$D bulk with coordinates $(x^\mu, y)$ and metric $G_{AB}$. For simplicity, take a flat bulk with
\begin{equation}
    ds^2 = -v_{\rm h}^2 \, dt^2 + d\mathbf{x}^2 + dy^2,
    \label{eq:bulk_metric}
\end{equation}
where $v_{\rm h} > c$ is the fundamental signal speed for $\Phi$.

A minimal bulk action is
\begin{equation}
    S_\Phi = \int d^4x \, dy \, \sqrt{-G}
    \left[
        -\frac{1}{2} G^{AB} \partial_A \Phi \, \partial_B \Phi - V(\Phi)
    \right],
\end{equation}
augmented by a localized interaction with the SM stress-energy trace:
\begin{equation}
    S_{\rm int}^{(1)} = \int d^4x \, dy \, \sqrt{-G} \;
    g_1 \, \Phi(x,y) \, T(x) \, \delta(y),
    \label{eq:layer1_int}
\end{equation}
where $g_1$ is a coupling constant of mass dimension $-1$ (for a canonically normalized $\Phi$).

Varying with respect to $\Phi$ yields the bulk equation of motion
\begin{equation}
    \left(
        -\frac{1}{v_{\rm h}^2} \frac{\partial^2}{\partial t^2}
        + \nabla^2
        + \frac{\partial^2}{\partial y^2}
    \right) \Phi(x,y) + V'(\Phi)
    = g_1 \, T(x) \, \delta(y).
    \label{eq:Phi_eom}
\end{equation}
Neglecting $V'(\Phi)$ for small perturbations, a nuclear detonation with localized $T(x)$ on the brane ($y = 0$) acts as a source for $\Phi$ modes that propagate in the bulk with speed $v_{\rm h}$.

From the brane viewpoint, the Green's function $G_{\rm R}(x^\mu, 0; x'^\mu, 0)$ of the operator in Eq.~\eqref{eq:Phi_eom} has support inside the $v_{\rm h}$-light cone, which is wider than the SM light cone defined by $c$. Hence, $\Phi$-mediated disturbances can correlate spacetime points that are spacelike with respect to $g_{\mu\nu}$, while remaining causal in the bulk metric $G_{AB}$.

For comparison with other portal scenarios, we can also phrase the interaction purely in 4D effective terms. Dimensional consistency requires the operator to have mass dimension 4. For a scalar $\Phi$ (dimension 1) and trace $T$ (dimension 4), the coefficient must have dimension $-1$. We thus write:
\begin{equation}
    \mathcal{L}_{\rm eff}^{(1)} \sim \frac{c_{\Phi}}{\Lambda} \, T(x) \, \Phi(x,0),
    \label{eq:L_eff_dim5}
\end{equation}
where $\Lambda$ represents the effective scale of the portal (related to the bulk scale and warping) and $c_{\Phi}$ is a dimensionless coefficient. The equation of motion for $\Phi$ then takes the form
\begin{equation}
    \Box \Phi + m_\Phi^2 \Phi = \frac{c_{\Phi}}{\Lambda} \, T(x),
    \label{eq:Phi_4d_eom}
\end{equation}
so that the detonation provides a rapidly varying source. The source is largest during the microsecond timescale of the burn front, when $T(x)$ changes most dramatically.

\subsection{Layer~0: Substrate excitations and trans-metric shocks}
\label{subsec:layer0_coupling}

At the substrate level (Layer~0), we imagine an order parameter field $\Sigma$ whose vacuum expectation value $\langle \Sigma \rangle$ encodes the emergent metric $g_{\mu\nu}$. In analogy to superfluid or condensed-matter systems, perturbations of $\Sigma$ propagate with a microphysical velocity $V_{\rm sub}$ that sets the true causal structure.

Let $\delta\Sigma$ denote small deviations from equilibrium. A schematic equation of motion is
\begin{equation}
    \Box_{\rm sub} \, \delta\Sigma
    + U'(\delta\Sigma)
    = \frac{1}{\Lambda^2} \mathcal{S}[T^{\mu\nu}],
    \label{eq:sigma_eom}
\end{equation}
where $\Box_{\rm sub}$ is the d'Alembertian associated with the substrate dynamics (characteristic speed $V_{\rm sub}$), $U$ an effective potential, $\Lambda$ an emergent symmetry-breaking scale, and $\mathcal{S}[T^{\mu\nu}]$ some functional of the SM stress-energy. A simple illustrative choice is
\begin{equation}
    \Box_{\rm sub} \, \delta\Sigma
    + m_\Sigma^2 \, \delta\Sigma
    = \frac{\xi}{\Lambda^2} \,
    (\partial_\mu T^{\nu\rho}) (\partial^\mu T_{\nu\rho}),
    \label{eq:sigma_source}
\end{equation}
where $\xi$ is dimensionless. Unlike $\partial_\mu T^{\mu\nu}$, which vanishes for conserved stress-energy, the combination $(\partial_\mu T^{\nu\rho})(\partial^\mu T_{\nu\rho})$ remains nonzero whenever $T^{\mu\nu}$ varies in spacetime. This source term is negligible in equilibrium but becomes significant in rapidly evolving configurations such as nuclear detonations, where all components of $T^{\mu\nu}$ change dramatically on short timescales.

In this ``trans-metric shockwave'' picture, gravitational waves are ripples \emph{of} the emergent metric $g_{\mu\nu}$ (Layer~2), while $\delta\Sigma$ excitations are ripples \emph{in} the substrate that generates $g_{\mu\nu}$ (Layer~0). Their dispersion relation might take the form
\begin{equation}
    \omega^2 = V_{\rm sub}^2 k^2 + m_\Sigma^2,
    \label{eq:sigma_dispersion}
\end{equation}
with $V_{\rm sub} \gg c$. Such waves would be invisible to detectors coupled only to $g_{\mu\nu}$ and SM fields, but in principle they could reflect global disruptions of the underlying ``spacetime fluid.''

\subsection{Other portal mechanisms}

Beyond stress-energy trace and substrate couplings, one can consider additional portal interactions that nuclear detonations might activate:
\begin{itemize}
    \item \textit{Curvature portal:}
    \begin{equation}
        \mathcal{L} \supset \xi_R R \, \Phi^2,
    \end{equation}
    with $R$ the Ricci scalar of $g_{\mu\nu}$ and $\xi_R$ a dimensionless parameter. Local curvature enhancements near dense matter could source $\Phi$.
    \item \textit{Gauge kinetic mixing:}
    \begin{equation}
        \mathcal{L} \supset \frac{\epsilon}{2} B^{\mu\nu} X_{\mu\nu},
    \end{equation}
    where $B^{\mu\nu}$ is the SM hypercharge field strength and $X_{\mu\nu}$ a hidden $U(1)$ field. Rapid changes in EM fields during a detonation could excite $X$.
    \item \textit{Neutrino portal:}
    Hidden fermions $\chi$ could couple via
    \begin{equation}
        \mathcal{L} \supset \frac{1}{\Lambda^2}
        (\bar{L} H)(\bar{\chi} \chi),
    \end{equation}
    where $L$ is a lepton doublet and $H$ the Higgs. The copious neutrino flux from nuclear processes might then source $\chi$.
\end{itemize}
These can all be embedded consistently within the Layer~1/Layer~0 framework, though we will keep our focus on stress-energy and substrate couplings for clarity.

\section{Mechanisms for apparent superluminality}
\label{sec:superluminal}

We now catalog several mechanisms by which the hidden and substrate layers can give rise to \emph{apparent} superluminal propagation when viewed from the emergent SM perspective.

\subsection{Extra-dimensional shortcuts}
\label{subsec:shortcuts}

In brane-world scenarios~\cite{ADD,RS1}, SM fields are confined to a 3+1-dimensional brane embedded in a higher-dimensional bulk. Bulk fields such as $\Phi$ can propagate in all dimensions, potentially taking geodesics that are shorter (in proper time) than any path confined to the brane.

Consider a simple warped 5D metric
\begin{equation}
    ds^2 = e^{-2k|y|} \eta_{\mu\nu} dx^\mu dx^\nu + dy^2,
\end{equation}
with $\eta_{\mu\nu}$ the Minkowski metric and $k$ a curvature scale. Two events separated by distance $L$ along the brane at $y = 0$ can be connected by a bulk null geodesic that dips into the $y$ direction. The proper path length $d_{\rm bulk}$ between these events satisfies
\begin{equation}
    d_{\rm bulk} < L,
\end{equation}
yielding an effective brane-projected velocity
\begin{equation}
    v_{\rm eff} = c \frac{L}{d_{\rm bulk}} > c.
\end{equation}
This is entirely compatible with the 5D causal structure and does not violate fundamental Lorentz invariance; the apparent superluminality is a projection effect.

In our notation, the bulk light cone is governed by $v_{\rm h}$ (or effectively $c$ in 5D units), while the emergent brane light cone by $c$ in 4D. Nuclear detonations provide intense boundary sources for $\Phi$, which can then traverse bulk shortcuts before re-interacting with the brane at distant locations.

\subsection{Modified dispersion relations}
\label{subsec:mdr}

Hidden-sector fields might obey modified dispersion relations,
\begin{equation}
    E^2 = p^2 c^2 + m^2 c^4
        + \alpha \frac{p^4}{M_P^2}
        + \beta \frac{p^6}{M_P^4} + \dots,
    \label{eq:mdr}
\end{equation}
where $M_P$ is the Planck scale (or another high scale), and $\alpha$, $\beta$ are dimensionless coefficients~\cite{AmelinoCameliaLRR}. The group velocity is
\begin{equation}
    v_g = \frac{\partial E}{\partial p}
    \approx c \left[
    1 + \alpha \frac{E^2}{M_P^2 c^4} + \dots
    \right].
\end{equation}
For $\alpha > 0$ and high energies, one can have $v_g > c$. Although $E / M_P c^2$ is tiny even for the most energetic particles we can produce, collective and coherent effects in large, rapidly varying field configurations might effectively amplify these corrections in an emergent description.

\subsection{Tachyonic hidden fields}

Tachyonic fields with $m^2 < 0$ have dispersion
\begin{equation}
    E^2 = p^2 c^2 - |m|^2 c^4,
\end{equation}
leading to group velocities
\begin{equation}
    v_g = \frac{p c^2}{E} =
    \frac{c}{\sqrt{1 - |m|^2 c^4 / E^2}} > c.
\end{equation}
In conventional quantum field theory, such modes signal vacuum instability; the system typically condenses into a new vacuum in which excitations are ordinary, $m^2 > 0$ particles.

In hidden sectors, however, tachyonic dispersion can appear in more subtle ways, e.g., as an effective description of localized excitations or in extra-dimensional setups with nontrivial boundary conditions~\cite{Feinberg1967}. If a nuclear detonation can transiently push the system into a regime where tachyonic modes are excited, these could mediate superluminal signals before decaying or relaxing back.

Any coupling of such tachyonic modes to SM operators,
\begin{equation}
    \mathcal{L}_{\rm int} \sim
    \frac{g_t}{\Lambda^n} \,
    \mathcal{O}_{\rm SM} \mathcal{O}_{\rm tach},
\end{equation}
would need to be extremely weak to avoid catastrophic instabilities in observed physics.

\subsection{Substrate waves}
\label{subsec:substrate_waves}

At the deepest layer, substrate excitations described by Eqs.~\eqref{eq:sigma_eom}--\eqref{eq:sigma_dispersion} can propagate at $V_{\rm sub} \gg c$. Because $g_{\mu\nu}$ and SM fields are emergent from $\Sigma$, these waves are not constrained by the emergent light cone; they are constrained by the microphysical causal cone.

From an SM perspective, substrate waves would look like acausal, nonlocal disturbances in effective spacetime or in the quantum state of the fields. In practice, detection would require either:
\begin{enumerate}
    \item a non-negligible backreaction of $\delta\Sigma$ on $g_{\mu\nu}$ or SM interactions, beyond the usual weak-gravity limit; or
    \item engineered detectors that are sensitive directly to $\Sigma$-dependent quantities, analogous to how certain condensed-matter probes access order-parameter fluctuations.
\end{enumerate}

This is analogous to how, in a crystal, the rearrangement of electron clouds at EM speed $c$ (deep layer) can influence phonons (emergent layer) through changes in interatomic potentials, even though the elastic theory alone would not predict such fast effects.

\section{Nuclear detonations as exceptional probes}
\label{sec:nuclear}

\subsection{Scales and nonequilibrium measures}

Within the stratified framework, nuclear detonations stand out because they simultaneously maximize several measures of departure from equilibrium:
\begin{itemize}
    \item Rapid changes in $T^{\mu\nu}$ on microsecond timescales;
    \item Large energy density gradients $\nabla \rho_E$ on sub-meter scales;
    \item High entropy production rate $\dot{S}$.
\end{itemize}
One can schematically write an effective coupling strength between SM and hidden/substrate sectors as
\begin{equation}
    g_{\rm eff} \sim g_0 \,
    f\!\left(\frac{\dot{S}}{k_B}\right)
    h(R_{\mu\nu\rho\sigma})
    \, \Gamma(\partial_t T^{\mu\nu}),
\end{equation}
where $f$, $h$, and $\Gamma$ are increasing functions of their arguments, and $R_{\mu\nu\rho\sigma}$ is the Riemann tensor. In ordinary laboratory conditions, $g_{\rm eff}$ would be negligible; in nuclear detonations, it might reach its largest terrestrial values.

For a rough estimate of entropy production,
\begin{equation}
    \dot{S} \sim \frac{E_{\rm total}}{T_{\rm eff} \, \tau},
\end{equation}
where $T_{\rm eff} \sim 10^7$--$10^8$~K is a characteristic temperature in the fireball. Plugging $E_{\rm total} \sim 10^{14}$~J and $\tau \sim 10^{-6}$~s suggests $\dot{S}$ values orders of magnitude above typical laboratory processes.

\subsection{Nonlinear regime and ``metric breaking''}

If the emergent metric $g_{\mu\nu}$ is itself an order parameter of the substrate, extreme local stress-energy might temporarily drive the system out of the linear-response regime in which general relativity holds. In such a region:
\begin{itemize}
    \item The relation between $T^{\mu\nu}$ and the Einstein tensor $G_{\mu\nu}$ may break down;
    \item $\delta\Sigma$ fluctuations may become large, corresponding to local ``phase slips'' of the spacetime condensate;
    \item The effective causal cone for SM fields might be deformed, while substrate and hidden-sector signals continue to propagate according to deeper microphysical laws.
\end{itemize}

We can think of this as an analogue of a region in a superfluid where the flow velocity exceeds the local sound speed: vortices, shocks, or phase defects can form, and the linear hydrodynamic description fails. In our case, the ``supersonic'' regime is in the space of stress-energy and curvature.

\section{Mini-shockwaves in particle accelerators}
\label{sec:accelerators}

\subsection{Accelerator collisions as localized stress-energy spikes}

While nuclear detonations provide the largest integrated energy release among controlled terrestrial events, particle accelerator collisions offer a complementary regime: extremely high peak energy densities confined to femtometer-scale volumes and attosecond-scale durations. In high-energy hadron collisions at facilities such as the Large Hadron Collider (LHC), center-of-mass energies of $\sqrt{s} \sim 10$~TeV are deposited into interaction volumes of order
\begin{equation}
    V_{\rm coll} \sim (1~{\rm fm})^3 \sim 10^{-45}~{\rm m}^3,
\end{equation}
over timescales
\begin{equation}
    \tau_{\rm coll} \sim \frac{1~{\rm fm}}{c} \sim 10^{-24}~{\rm s}.
\end{equation}

The resulting instantaneous energy density can be enormous:
\begin{equation}
    \rho_E^{\rm coll} \sim \frac{E_{\rm cm}}{V_{\rm coll}}
    \sim \frac{10^{4}~{\rm GeV}}{10^{-45}~{\rm m}^3}
    \sim 10^{36}~{\rm J/m}^3,
\end{equation}
exceeding nuclear detonation energy densities by over twenty orders of magnitude. These collisions thus represent ``mini-shockwaves'' in the stress-energy tensor: extremely peaky perturbations with negligible integrated energy but unparalleled local intensity.

\subsection{Comparison of source characteristics}

\begin{table}[h]
\caption{Comparison of nuclear detonations and accelerator collisions as hidden-sector sources.}
\label{tab:source_comparison}
\begin{ruledtabular}
\begin{tabular}{lrr}
Parameter & Nuclear   & Accelerator  \\
 &  detonation &  collision \\
\hline
Total energy $E_{\rm tot}$ & $10^{14}$~J & $10^{-6}$~J \\
Duration $\tau$ & $10^{-6}$~s & $10^{-24}$~s \\
Spatial scale $L$ & $1$~m & $10^{-15}$~m \\
Peak energy density $\rho_E$ & $10^{14}$~J/m$^3$ & $10^{36}$~J/m$^3$ \\
Peak power $P$ & $10^{20}$~W & $10^{18}$~W \\
$|\partial_t T^{\mu\nu}|$ & Large & Extreme \\
Coherence & Macroscopic & Microscopic \\
Repetition rate & Rare & $10^{7}$--$10^{9}$/s \\
\end{tabular}
\end{ruledtabular}
\end{table}

Table~\ref{tab:source_comparison} summarizes the key differences between nuclear detonations and accelerator collisions as potential sources of hidden-sector or substrate excitations.
The two source types occupy opposite corners of the parameter space: nuclear detonations maximize integrated energy and macroscopic coherence, while accelerator collisions maximize peak density and temporal sharpness.

\subsection{Coupling efficiency for peaky sources}

Within our effective field theory framework, the relative efficiency of these sources depends on which coupling operators dominate. For the trace coupling of Eq.~\eqref{eq:L_eff_dim5},
\begin{equation}
    \mathcal{L}_{\rm eff}^{(1)} \sim \frac{c_{\Phi}}{\Lambda} \, T(x) \, \Phi,
\end{equation}
the amplitude of hidden-sector excitation scales with the spacetime integral of the source:
\begin{equation}
    \Phi_{\rm amp}^{(1)} \propto \int d^4x \, T(x) \propto E_{\rm tot}.
\end{equation}
In this case, nuclear detonations dominate by a factor of $\sim 10^{20}$ over individual accelerator collisions.

However, for gradient couplings sensitive to rapid temporal variation, such as Eq.~\eqref{eq:sigma_source},
\begin{equation}
    \mathcal{S}[T^{\mu\nu}] \propto (\partial_\mu T^{\nu\rho})(\partial^\mu T_{\nu\rho}),
\end{equation}
the source strength scales as
\begin{equation}
    \mathcal{S} \propto \left(\frac{E_{\rm tot}}{\tau L^3}\right)^2 \tau L^3
    = \frac{E_{\rm tot}^2}{\tau L^3}.
\end{equation}
For this measure, accelerator collisions can compete with or exceed nuclear detonations despite their minuscule integrated energy:
\begin{equation}
    \frac{\mathcal{S}_{\rm coll}}{\mathcal{S}_{\rm nuc}}
    \sim \frac{E_{\rm coll}^2 \, \tau_{\rm nuc} \, L_{\rm nuc}^3}
              {E_{\rm nuc}^2 \, \tau_{\rm coll} \, L_{\rm coll}^3}
    \sim 10^{10} \text{--} 10^{20},
\end{equation}
depending strongly on the effective coherence length $L_{\rm coll}$ assigned to the parton-level interaction vertex.

Similarly, for nonlinear couplings proportional to $(T_{\mu\nu} T^{\mu\nu})^n$ with $n > 1$, the extreme local energy density of accelerator collisions provides strong enhancement:
\begin{equation}
    (T_{\mu\nu} T^{\mu\nu})_{\rm coll} \sim \rho_E^2 \sim 10^{72}~{\rm J}^2/{\rm m}^6,
\end{equation}
compared to $\sim 10^{28}~{\rm J}^2/{\rm m}^6$ for nuclear detonations.

\subsection{Collective effects and repetition}

A crucial distinction is the repetition rate. Modern colliders produce $\sim 10^{9}$ collisions per second, potentially enabling statistical accumulation of hidden-sector effects even if individual collision amplitudes are small. If each collision produces a substrate or hidden-sector excitation with probability $p$ and amplitude $a$, the cumulative signal after $N$ collisions scales as:
\begin{equation}
    \text{Coherent:} \; A_N \propto N a, \qquad
    \text{Incoherent:} \; A_N \propto \sqrt{N} \, a.
\end{equation}

For $N \sim 10^{17}$ collisions per year at the LHC, even incoherent accumulation provides an enhancement factor of $\sim 10^{8}$. Whether such accumulation is physically realized depends on the coherence properties of the hidden-sector excitations and the detection mechanism.

\subsection{Mini-shockwaves and substrate response}

In the trans-metric shockwave picture of Sec.~\ref{subsec:layer0_coupling}, individual accelerator collisions can be viewed as microscopic ``hammer blows'' on the spacetime substrate. Each collision produces a localized, delta-function-like perturbation:
\begin{equation}
    T^{\mu\nu}(x) \approx E_{\rm cm} \, u^\mu u^\nu \, \delta^{(3)}(\mathbf{x} - \mathbf{x}_0) \, \delta(t - t_0),
\end{equation}
where $u^\mu$ is the four-velocity of the center-of-mass frame.

The substrate response to such an impulsive source is determined by the Green's function of Eq.~\eqref{eq:sigma_eom}:
\begin{equation}
    \delta\Sigma(x) = \int d^4x' \, G_{\rm sub}(x; x') \, \mathcal{S}[T^{\mu\nu}(x')].
\end{equation}
For a delta-function source, this reduces to
\begin{equation}
    \delta\Sigma(x) \propto \frac{\mathcal{S}_0}{\Lambda^2} \, G_{\rm sub}(x; x_0),
\end{equation}
where $\mathcal{S}_0$ encodes the collision intensity and $G_{\rm sub}$ propagates the disturbance at velocity $V_{\rm sub}$.

The resulting substrate wave is a spherical pulse expanding at $V_{\rm sub} \gg c$, with amplitude determined by the local intensity of the collision. While the integrated energy in this pulse is negligible, its peak amplitude and the sharpness of its wavefront may be relevant for certain detection schemes sensitive to transient perturbations.

\subsection{Accelerator-based detection strategies}

The high repetition rate and controlled timing of accelerator collisions suggest specific detection strategies distinct from those applicable to nuclear detonations:

\begin{enumerate}
    \item \textit{Beam-dump experiments:} Detectors placed behind shielding that blocks all SM particles could search for anomalous energy deposition correlated with beam timing. Hidden-sector or substrate excitations produced in collisions might reconvert in dense material, depositing detectable energy.

    \item \textit{Missing energy/momentum:} Precision measurements of collision kinematics can reveal energy or momentum escaping into undetected channels. Current searches for dark matter and other hidden sectors already exploit this signature; the framework here suggests additional motivation for such analyses.

    \item \textit{Timing correlations:} If substrate excitations propagate at $V_{\rm sub} \gg c$, detectors at different distances from the interaction point might observe correlated signals with timing inconsistent with luminal propagation. The precisely known collision times at accelerators provide an ideal reference for such searches.

    \item \textit{Precision interferometry:} Accelerator facilities could host gravitational-wave-style interferometers searching for correlated noise bursts synchronized with collision events. The high repetition rate enables statistical accumulation of weak signals.
\end{enumerate}

\subsection{Complementarity of source types}

The existence of two qualitatively different source regimes---macroscopic/energetic (nuclear detonations) and microscopic/peaky (accelerator collisions)---provides important theoretical leverage. If hidden-sector or substrate couplings were observed in one regime but not the other, this would strongly constrain the form of the coupling operators.

For example:
\begin{itemize}
    \item Observation in nuclear detonations but not accelerators would favor couplings proportional to integrated energy or macroscopic coherence effects.
    \item Observation in accelerators but not nuclear detonations would favor gradient or nonlinear couplings sensitive to peak intensity rather than total energy.
    \item Observation in both, with predictable relative strengths, would provide a powerful consistency check on the underlying theory.
\end{itemize}

This complementarity mirrors the situation in condensed matter physics, where both macroscopic mechanical shocks and microscopic particle impacts can excite phonon and electronic degrees of freedom, with different efficiency depending on the coupling mechanism.

\subsection{Heavy-ion collisions and quark-gluon plasma}

A particularly interesting intermediate regime is provided by heavy-ion collisions at RHIC and the LHC. In central Pb-Pb or Au-Au collisions, thousands of nucleons participate, creating a quark-gluon plasma (QGP) with:
\begin{equation}
    E_{\rm tot}^{\rm HI} \sim 10^{3}~{\rm TeV} \sim 10^{-4}~{\rm J},
\end{equation}
over volumes $V \sim (10~{\rm fm})^3 \sim 10^{-42}~{\rm m}^3$ and durations $\tau \sim 10^{-23}$~s.

The QGP phase represents a genuine nonperturbative restructuring of the QCD vacuum. The trace of the stress-energy tensor, $T^\mu{}_\mu$, is dominated at low energies by the chiral condensate $\langle \bar{q} q \rangle$ via the trace anomaly. During the deconfinement transition, this condensate ``melts,'' leading to a large, non-perturbative excursion in $T^\mu{}_\mu$ that cannot be modeled by perturbative gluon bremsstrahlung alone. If the substrate $\Sigma$ or hidden sector couples to the conformal anomaly, heavy-ion collisions provide a unique pumping mechanism: a rapid, coherent volume-switching of the vacuum state itself, rather than a mere redistribution of particle momenta.

Heavy-ion collisions thus offer a ``bridge'' regime between proton-proton collisions (maximum peakiness) and nuclear detonations (maximum integrated energy), potentially probing coupling structures inaccessible to either extreme alone.

\section{Causality, consistency, and constraints}
\label{sec:constraints}

\subsection{Microcausality and emergent Lorentz invariance}

A central challenge is to reconcile apparently superluminal propagation with causality. In special relativity, controllable superluminal signals generically lead to closed timelike curves in some inertial frame.

Within our stratified picture, several escape routes exist:
\begin{enumerate}
    \item \textit{Fundamental causal structure at $V_{\rm sub}$:} Microcausality is defined with respect to the substrate dynamics and its causal speed $V_{\rm sub}$. Both the hidden sector (Layer~1) and SM (Layer~2) emerge from this deeper level. What appears superluminal with respect to $c$ may still be subluminal with respect to $V_{\rm sub}$.
    \item \textit{Emergent Lorentz invariance:} The SM sector can enjoy an approximate Lorentz invariance with invariant speed $c$ even if the underlying theory has a different invariant speed, so long as couplings that violate SM Lorentz symmetry are extremely small~\cite{LiberatiReview}.
    \item \textit{Instability of Causal Loops:} Consistent with Hawking's chronology protection conjecture, the backreaction of vacuum fluctuations might diverge on any emerging closed timelike curve, dynamically preventing the utilization of superluminal modes for macroscopic time travel.
    \item \textit{Preferred frame:} The hidden or substrate sectors might single out a preferred frame (e.g., the cosmological rest frame), breaking global Lorentz symmetry while preventing closed timelike curves~\cite{Hawking1992}.
\end{enumerate}

Maintaining consistency requires that radiative corrections from hidden/substrate sectors do not induce large Lorentz-violating terms in SM dispersion relations, which are tightly constrained by high-energy astrophysics and precision tests~\cite{LiberatiReview}.

\subsection{Energy-budget constraints}

Any energy emitted into hidden or substrate sectors must be compatible with measured energetics of nuclear reactions. Present data on fission and fusion yields are consistent with SM predictions at the percent level, implying
\begin{equation}
    \frac{E_{\rm hidden}}{E_{\rm total}} \lesssim 10^{-2}
\end{equation}
for terrestrial nuclear processes.

Far stronger constraints arise from astrophysics and cosmology. If ordinary nuclear burning in stars or core-collapse supernovae efficiently leaked energy into hidden sectors via couplings like Eq.~\eqref{eq:L_eff_dim5} or Eq.~\eqref{eq:sigma_source}, stellar structure, supernova light curves, and neutrino signals would be modified. For example, the observed neutrino burst from SN~1987A lasting $\sim 10$~s constrains additional cooling channels, while stellar lifetime measurements limit anomalous energy loss in the solar core. These considerations typically require portal strengths such as
\begin{equation}
    \frac{c_{\Phi}}{\Lambda} \ll 10^{-10}~{\rm GeV}^{-1},
\end{equation}
depending on the operator and mass scales involved~\cite{Raffelt}.

However, the terrestrial advantage lies in proximity and triggering. The flux of hidden-sector particles at a detector scales as $r^{-2}$; a detector located 1~km from a nuclear test intercepts a flux factor of $\sim 10^{32}$ larger than one observing a supernova at 50~kpc. Thus, even if nuclear detonations are inefficient producers, the ability to position detectors within the ``near field'' of the source offers a complementary discovery space.

\section{Observational signatures and detection strategies}
\label{sec:observables}

\subsection{Precursor signals and time-of-flight anomalies}

If nuclear explosions emit hidden or substrate signals that later reconvert into SM observables at distant locations, one generic signature would be \emph{precursor} events: detectable signals arriving earlier than any light-speed-limited signal could.

For a hidden-sector propagation speed $v_{\rm h} > c$ over distance $d$, the arrival-time advance is
\begin{equation}
    \Delta t = \frac{d}{c} - \frac{d}{v_{\rm h}}
    = \frac{d}{c} \left(1 - \frac{c}{v_{\rm h}}\right).
\end{equation}
For example, taking $v_{\rm h} = 2c$ and $d = 10^4$~km (Earth scale) yields
\begin{equation}
    \Delta t \approx 17~{\rm ms}.
\end{equation}
This is comfortably within the resolution of modern timing networks.

Similar arguments apply if substrate waves with speed $V_{\rm sub} \gg c$ can trigger detectable SM responses far away, though the mechanism of reconversion is more speculative.

\subsection{Reconversion mechanisms}

Detection requires that hidden or substrate excitations eventually produce SM quanta or metric perturbations in a way that can be measured. Possible reconversion channels include:
\begin{itemize}
    \item \textit{Inverse portal processes:} The same operators responsible for production (e.g., Eq.~\eqref{eq:L_eff_dim5}) can mediate $\Phi \Phi \to {\rm SM}$ annihilation in regions with nontrivial $T^{\mu\nu}$.
    \item \textit{Coherent conversion in backgrounds:} Analogous to axion--photon conversion in magnetic fields, a hidden scalar or vector might mix with photons in the presence of strong electromagnetic or gravitational backgrounds.
    \item \textit{Gravitational backreaction:} Substrate disturbances could induce small, nonstandard metric perturbations that appear as anomalous gravitational-wave-like signals or as low-level noise in precision interferometers.
\end{itemize}
All such mechanisms are highly model dependent and severely constrained by null results in analogous searches (e.g., axion-like particle experiments).

\subsection{Retrospective correlation studies}

Even without an explicit microphysical model, one can search statistically for anomalous correlations associated with historical nuclear tests:
\begin{enumerate}
    \item Cross-correlate the times and locations of past detonations with data from global networks (seismic, EM transient, neutrino detectors, cosmic-ray monitors) for excess coincidences at time offsets incompatible with luminal propagation.
    \item Look for directional patterns in putative anomalous signals that might hint at extra-dimensional geometries or substrate anisotropies.
    \item Search for energy- or frequency-dependent signatures aligning with generic expectations from Eqs.~\eqref{eq:mdr}, \eqref{eq:Phi_eom}, or \eqref{eq:sigma_dispersion}.
\end{enumerate}
Such analyses are challenging due to the complexity of natural backgrounds and the lack of sharp theoretical predictions, but they represent a way to leverage existing data.

\subsection{Dedicated experiments}

Future experiments, possibly using subcritical nuclear tests or large conventional explosions, could be designed to maximize sensitivity to hidden or substrate signals:
\begin{itemize}
    \item \textit{Deeply shielded detectors} located near the source, engineered to be insensitive to known particle and radiation backgrounds, but sensitive to unexpected energy depositions or timing anomalies.
    \item \textit{Global timing arrays} of high-precision clocks and detectors, synchronized at the microsecond or better level, searching for correlated, directionally reconstructible transients.
    \item \textit{Interferometric probes} (e.g., advanced gravitational-wave detectors) operating during controlled high-energy events, searching for coincident noise bursts or anomalies with statistically significant correlations to the source.
\end{itemize}
While such programs would be technically and politically nontrivial, they exemplify the kind of experimental approach that a speculative framework of this sort would motivate.

\section{Discussion}
\label{sec:discussion}

We have assembled a stratified, unified framework for various speculative ideas about superluminal signaling sourced by nuclear detonations. The main conceptual ingredients are:
\begin{itemize}
    \item An emergent view of spacetime and the SM, inspired by analogue-gravity models, in which our familiar light speed $c$ is analogous to a sound speed $v_s$ rather than a fundamental constant;
    \item Hidden-sector fields $\Phi$ living in higher-dimensional or otherwise extended spacetimes, which can exploit geometric shortcuts or modified dispersion relations to appear superluminal in 3+1 dimensions;
    \item A deeper substrate $\Sigma$ whose excitations propagate at $V_{\rm sub} \gg c$, with the emergent metric $g_{\mu\nu}$ representing a coarse-grained order parameter;
    \item Nuclear detonations as intense, rapidly varying sources of stress-energy that can, in principle, couple efficiently (though still feebly in absolute terms) to both $\Phi$ and $\Sigma$.
\end{itemize}

The acoustic--electromagnetic analogy plays a central role: it shows concretely that hierarchies of characteristic speeds and cross-phenomenal couplings are common in effective theories, and that emergent relativistic structures can coexist with deeper, faster microphysics. In solids, phonons and photons share a common atomic substrate; here, we extend the analogy by hypothesizing a shared substrate underlying both SM and hidden sectors.

The framework is heavily constrained by established physics:
\begin{itemize}
    \item Lorentz invariance tests severely limit any observable violations or superluminal propagation in the SM sector;
    \item Astrophysical and cosmological observations constrain additional energy-loss channels from known astrophysical engines;
    \item Precision calorimetry and reaction modeling constrain exotic channels in terrestrial nuclear processes.
\end{itemize}
These considerations strongly suggest that, if hidden or substrate sectors of the type described here exist, their couplings to the SM are extremely small, and any associated signals from nuclear detonations would be correspondingly tiny.

Nevertheless, the conceptual payoff is twofold. First, the exercise sharpens our understanding of how emergent relativity and deeper causal structures might coexist, and what sorts of phenomena (e.g., apparent superluminality) they could, in principle, accommodate. Second, it identifies extreme, controlled events such as nuclear detonations as natural laboratories in which to \emph{search} for unexpected couplings, in the same way that laboratory experiments revealed acousto-optic effects in materials before they were theoretically understood.

\section{Conclusions}
\label{sec:conclusion}

We have presented a stratified framework that attempts to unify various speculative scenarios regarding superluminal signaling and emergent spacetime. By treating the speed of light $c$ as an emergent characteristic rather than a fundamental limit---analogous to the speed of sound in condensed matter---we categorize potential phenomenological extensions into hidden-sector fields (Layer~1) and deeper substrate excitations (Layer~0).

A central feature of this proposal is the identification of nuclear detonations as unique, macroscopic probes of the vacuum structure. While particle accelerators achieve higher peak energy densities, nuclear detonations operate in a distinct regime characterized by:
\begin{enumerate}
    \item \textbf{Macroscopic coherence:} The stress-energy tensor $T^{\mu\nu}$ is perturbed nonlinearly over volume scales ($V \sim 1~{\rm m}^3$) and timescales ($\tau \sim 1~\mu{\rm s}$) that are vast compared to fundamental interaction times. This allows for the potential excitation of collective, low-frequency substrate modes that point-like collisions cannot access.
    \item \textbf{Trace dynamics:} The violent evolution of the QCD trace anomaly and hydrodynamic stress during a detonation maximizes the scalar coupling $\int d^4x \, T(x)$ to hidden sectors, a quantity that is negligible in microscopic scattering events.
    \item \textbf{Nonequilibrium thermodynamics:} The extreme entropy production and ``metric breaking'' conditions near the fireball may push the local vacuum state outside the linear-response regime of general relativity, potentially unshielding the deeper causal structure of the substrate.
\end{enumerate}
Consequently, if hidden sectors couple to the \emph{bulk} properties of matter or spacetime curvature, nuclear tests represent the most luminous terrestrial sources available for such searches.

We contrasted these macroscopic sources with high-energy particle accelerators, which act as microscopic ``mini-shockwaves.'' Accelerators probe the complementary high-gradient, high-frequency limit. If couplings depend on $\partial_\mu T^{\nu\rho}$ or local energy density squared, accelerator collisions would dominate over nuclear sources. A consistent theory of trans-metric physics must account for why anomalies might appear in one regime but not the other, providing a powerful discriminator between coupling mechanisms (e.g., coherent vs.\ incoherent excitation).

It must be emphasized that this framework is constrained severely by existing data. The absence of gross Lorentz violations or energy-loss anomalies in astrophysics implies that any coupling between the Standard Model and these hidden layers is exceedingly weak. However, the acoustic analogy reminds us that emergent sectors can be robustly insulated from their substrates.

While purely speculative, this unified formalism suggests that historical data from nuclear tests and modern data from the LHC could be reanalyzed for nonstandard correlations---specifically, precursor signals or time-of-flight anomalies that violate the emergent light speed $c$ while respecting the underlying microcausality $V_{\rm sub}$. Whether nature utilizes this freedom remains an open question, but the stratified approach provides the necessary language to ask it.

\begin{acknowledgments}
This text was partially created and revised with assistance from one or more of the following large language models: Gemini 3 Pro, claude-opus-4-5-20251101, and GPT-5.1-high. All content, ideas, and prompts were provided by the author.

This research was funded in whole or in part by the Austrian Science Fund (FWF) Grant \href{https://doi.org/10.55776/PIN5424624}{10.55776/PIN5424624}.
The author acknowledges TU Wien Bibliothek for financial support through its Open Access Funding Programme.

The author declares no conflict of interest.
\end{acknowledgments}

\bibliography{svozil}

\end{document}